\documentclass[a4paper,12pt]{article}
\usepackage{epsfig}
\usepackage{amssymb}
\usepackage{amsfonts}
\usepackage{graphics}
\usepackage{color,amsmath,bm}

\newskip\humongous \humongous=0pt plus 1000pt minus 1000pt

\newif\ifdtup

\catcode`\@=11

\@addtoreset{equation}{section}

\def\@normalsize{\@setsize\normalsize{15pt}\xiipt\@xiipt
\abovedisplayskip 14pt plus3pt minus3pt%
\belowdisplayskip \abovedisplayskip
\abovedisplayshortskip \z@ plus3pt%
\belowdisplayshortskip 7pt plus3.5pt minus0pt}

\def\small{\@setsize\small{13.6pt}\xipt\@xipt
\abovedisplayskip 13pt plus3pt minus3pt%
\belowdisplayskip \abovedisplayskip
\abovedisplayshortskip \z@ plus3pt%
\belowdisplayshortskip 7pt plus3.5pt minus0pt
\def\@listi{\parsep 4.5pt plus 2pt minus 1pt
     \itemsep \parsep
     \topsep 9pt plus 3pt minus 3pt}}

\relax

\catcode`@=12

\catcode`\@=11

\def\section{\@startsection{section}{1}{\z@}{3.5ex plus 1ex minus
   .2ex}{2.3ex plus .2ex}{\large\bf}}

\def\thesection{\arabic{section}}

\def\appendix{\setcounter{section}{0}
\def\thesection{\Alph{section}}}

\begin{document}

\newcommand{\beq}{\begin{equation}}
\newcommand{\eeq}{\end{equation}}
\newcommand{\bea}{\begin{eqnarray}}
\newcommand{\eea}{\end{eqnarray}}
\newcommand{\beas}{\begin{eqnarray*}}
\newcommand{\eeas}{\end{eqnarray*}}
\newcommand{\defi}{\stackrel{\rm def}{=}}
\newcommand{\non}{\nonumber}
\newcommand{\bquo}{\begin{quote}}
\newcommand{\enqu}{\end{quote}}
\newcommand{\mat}{\mathbf}
\def\d{\mathrm{d}}
\def\de{\partial}
\def\Tr{ \hbox{\rm Tr}}
\def\const{\hbox {\rm const.}}
\def\o{\over}
\def\im{\hbox{\rm Im}}
\def\re{\hbox{\rm Re}}
\def\bra{\langle}\def\ket{\rangle}
\def\Arg{\hbox {\rm Arg}}
\def\Re{\hbox {\rm Re}}
\def\Im{\hbox {\rm Im}}
\def\diag{\hbox{\rm diag}}
\def\longvert{{\rule[-2mm]{0.1mm}{7mm}}\,}
\def\Z{\mathbb Z}
\def\N{{\cal N}}
\def\tq{{\widetilde q}}
\def\W{{\cal W}}
\def\tQ{{\widetilde Q}}
\def\dag{{}^{\dagger}}
\def\p{{}^{\,\prime}}
\def\a{\alpha}
\def\Tr{ \hbox{\rm Tr}}
\def\tM{{\widetilde M}}
\def\tm{{\widetilde m}}
\def\T{{\cal T}}
\def\t{T}
\def\J{{\cal J}}
\def\V{{\sf V}}
\def\lcm{\mathrm{lcm}}

\begin{titlepage}
\begin{flushright}
hep-th/0507273\\
\end{flushright}

\bigskip

\begin{center}
{\Large

{\bf Domain Walls and Flux Tubes }
 }
\end{center}

\renewcommand{\thefootnote}{\fnsymbol{footnote}}
\bigskip
\begin{center}
{\large   Stefano Bolognesi }
 \vskip 0.20cm
\end{center}

\begin{center}
{\it      \footnotesize
Scuola Normale Superiore - Pisa, Piazza dei Cavalieri 7, Pisa, Italy \\
\vskip 0.10cm
and\\
\vskip 0.10cm
Istituto Nazionale di Fisica Nucleare -- Sezione di Pisa, \\
Via Buonarroti 2, Ed. C, 56127 Pisa,  Italy   \\  } \vskip 0.15cm
s.bolognesi@sns.it\\
\end {center}

\setcounter{footnote}{0}

\bigskip
\bigskip

\noindent
\begin{center} {\bf Abstract} \end{center}

We present a new vortex solution made of a domain wall
compactified into a cylinder and stabilized by the magnetic flux
within. When the thickness of the wall is much less than the
radius of the vortex some precise results can be obtained, such as
the tension spectrum and profile functions. This vortex can
naturally end on the wall that has created it, making the simplest
junction between a wall and a vortex. We then classify every kind
of junction between a flux tube and domain wall. The criteria for
classification are as follows: a flux can or can not end on the
wall, and when it ends, the flux  must go somewhere. Various
examples are discussed, including abelian and nonabelian theories,
as well as supersymmetric and non-supersymmetric theories.

\vfill

\begin{flushleft}
July, 2005
\end{flushleft}

\end{titlepage}

\bigskip

\hfill{}

\tableofcontents

\section{Introduction}

We present a new kind of vortex which is formed by a wall. It
occurs whenever a theory has two degenerate vacua, one in the
Coulomb phase, and one in the Higgs phase. The ordinary vortex in
the Higgs vacuum \cite{A,NO} can be thought of as the domain wall
interpolating between the two vacua compactified into a cylinder.
The rolled wall is stabilized by the magnetic field
inside.\footnote{This is reminiscent of the relation between
$Q$-kinks and $Q$-lumps discussed in \cite{Abraham:1992qv}.}

We then study the extent to which the thickness of the wall
$\Delta_W$ is much less than the radius of the vortex $R_V$. This
limit is reached for a large number of quanta of the magnetic
flux. Within these limits the spectrum of vortices is simple to
compute by means of classical arguments.  They are particular kind
of type I superconducting vortices with a tension that scales like
$n^{2/3}$. In addition, the profile functions of the vortex are
exactly computable within this limit.

This kind of flux tube naturally ends on the wall that has created
it. When a wall vortex ends on its wall, the magnetic flux it
carries is spread radially through the Coulomb vacuum on the other
side of the wall. The point on the wall at which the vortex ends
is seen from the Coulomb vacuum as a monopole of double charge.
The junction between the wall and the wall vortex can be thought
of as the final stage of a process in which a monopole in the
Coulomb phase is pushed against the wall.

This kind of wall vortex arises  in the nonabelian gauge theories
every time a domain wall interpolates between two vacua with
different confinement indices.

We then classify the junctions between domain walls and flux
tubes. The  previously discussed junction is only one particular
variety, and perhaps the most natural.  The two basic criteria for
distinguish junctions are:
\begin{itemize}
\item The tube can or can not end on the wall. \item  Where the
flux goes.
\end{itemize}

When the vortex ends on the wall and spread its flux across the
$2$-dimensional surface of the wall is where we find a D-brane
junction. This example has been widely studied in recent years
\cite{Gauntlett:2000de,Shifman:2002jm,nahiggs}. The driving force
behind these studies was, in fact, to find some field theoretical
analogues of the D-brane physics in string theory. This variety of
wall, in order to completely resemble a D-brane, must also
supports massless gauge fields in the effective low-energy action.

Another type of junction is found when the string ends on the wall
and the flux is captured by a particle confined within the wall.
In this case, there is no localization of the gauge field inside
the wall. The first example has been found in \cite{Witten:1997ep}
where it was shown that the $\mathbb{Z}_N$ strings of pure $SU(N)$
$\N=1$ super Yang-Mills can end on the domain wall that
interpolates between two chirally adjacent vacua.

The last type is a cross-junction, which is characterized by the
tube, rather than ending on the wall, crossing it instead and
becoming another flux tube in the opposite vacuum.  We can find
examples of it in pure $SU(N)$ $\N=1$ SYM when the domain wall
interpolates between two vacua whose phase shift has a common
divisor with $N$.

Finally we put our classification to the test in a particular
theory. In $\N=2$ pure SYM broken to $\N=1$ by a generic
superpotential for the adjoint field, there are a lot of vacua
with different confinement index \cite{Cachazo:2002zk}. We provide
examples in which all three types  of junctions are presented
simultaneously.

The paper is divided into two parts.  Section \ref{made} is
devoted to the study of the  wall vortex. In  \ref{abelian} we
consider the abelian gauge theory and its supersymmetric
extension. In \ref{naexistence} we study the wall vortex in
nonabelian gauge theories. In  Section \ref{classification} we
classify the domain wall/flux tube junctions. In \ref{local} we
consider a flux tube which ends on the wall, and in
\ref{crossubsection}, a flux tube which crosses the wall. Finally,
in  \ref{super} we give an example in which every junction studied
in the paper is presented simultaneously. In Appendix
\ref{yappendix} we provide a mechanism to obtain a flux tube/flux
tube junction, and we use it to build a Y junction for baryons in
$SU(3)$.

\section{The Vortex Formed by a Wall \label{made}}

\begin{figure}[h!t]
\begin{center}
\leavevmode \epsfxsize 11 cm \epsffile{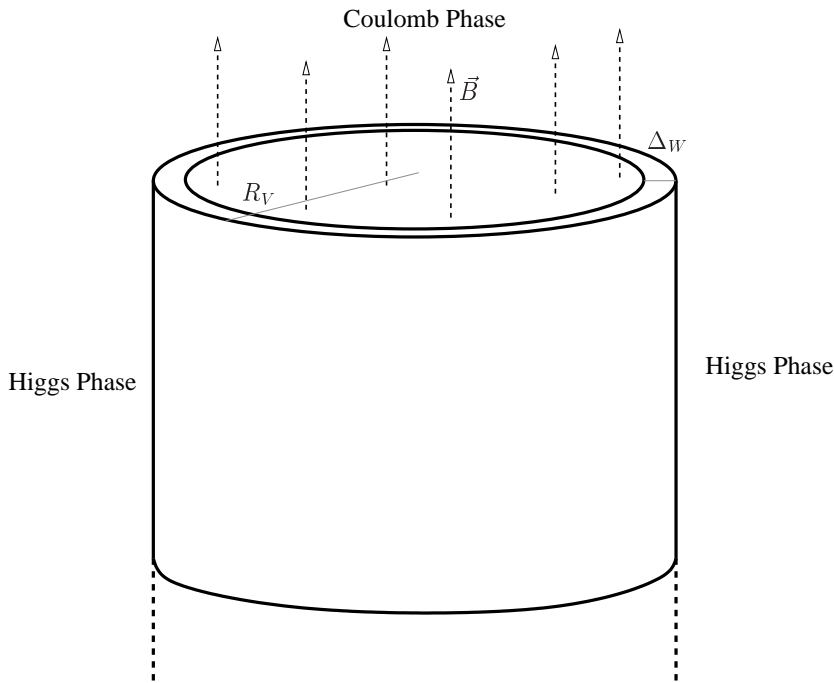}
\end{center}
\caption{\footnotesize The wall vortex. A wall of thickness
$\Delta_W$ is compactified on a circle of radius $R_V$ and
stabilized by the magnetic field inside.} \label{springroll}
\end{figure}

We consider a flux tube that can be thought as made of wall. The
simplest example in which it can arise is a $U(1)$ gauge theory
that has two degenerate vacua: one in the Coulomb phase and the
other in the Higgs phase. Consider the domain wall of tension
$T_W$ that interpolates between these two vacua. We can build a
flux tube rolling the wall in a cylinder of radius $R$, keeping
the Coulomb phase  inside the tube and turning on a magnetic flux
inside (see Figure \ref{springroll}).  The tension of the wall
gives a contribution $T_W 2\pi R$ to the energy density. The
magnetic flux is $\Phi_B=B \pi R^2$ where $B$ is the magnetic
field. Varying the radius $R$ the flux $\Phi_B$ must remain
constant, so the contribution of the flux to the energy density is
$\Phi_B / 2\pi R^2$.  The magnetic flux is quantized in integer
values: $\Phi_B=2\pi n/e$, where $e$ is the coupling constant. The
tension of the tube is the sum of two pieces, one that comes from
the flux and the other that comes from the wall: \beq \label{qual}
T(R)=\frac{{\Phi_B}^2}{2 \pi R^2}+T_W 2 \pi R \ .\eeq  The stable
configuration is the one that minimizes the tension: \beq
\label{eqscal} R_V = \sqrt[3]{2} \; \sqrt[3]{\frac{n^{2}}{ e^{2}
{T_W}}} \ , \qquad T_V=3 \sqrt[3]{2} \pi \; \left(\frac{n
T_W}{e}\right)^{2/3}  \ .\eeq This result can be trusted only when
the thickness of the wall $\Delta_W$ is much less than the radius
of the vortex.  In general the profile of the vortex will be a
mixture of the magnetic field and the wall. In any case this kind
of flux tube can be though as made of the same stuff of the wall.

\subsection{Abelian Higgs-Coulomb model \label{abelian}}

\begin{figure}[h!t]
\begin{center}
\leavevmode \epsfxsize 9 cm \epsffile{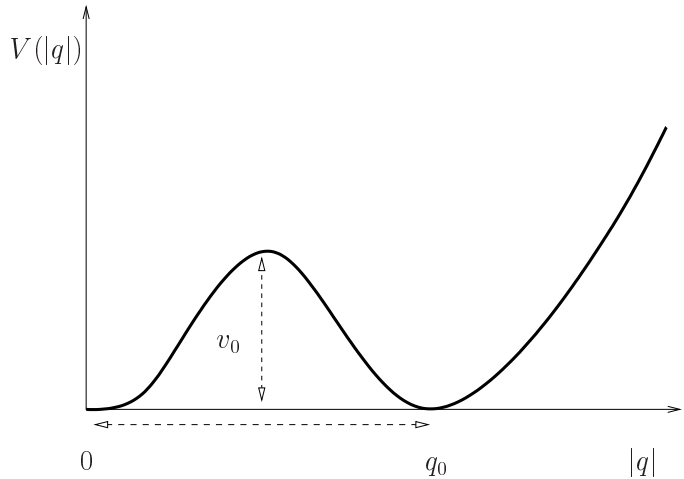}
\end{center}
\caption{\footnotesize A potential with two degenerate vacua.
$q=0$ is in the Coulomb phase while $|q|=q_0$ is in the Higgs
phase.} \label{potential}
\end{figure}

Now we are going to consider an explicit example. The simplest
that we can imagine is a $U(1)$ gauge theory coupled to a charged
scalar $q$
\begin{equation}
\label{BasicVortex} {\cal
L}=-\frac{1}{4}F_{\mu\nu}F^{\mu\nu}-\frac12
|(\de_{\mu}-ieA_{\mu})q|^2-V(|q|)\ ,
\end{equation}
and a suitable potential (see Figure \ref{potential}) that has two
degenerate  minima, one in the Coulomb phase $q=0$ and the other
in the Higgs phase $|q|=q_0$. 
This theory admits a kink that interpolates between the two vacua.
The profile of the kink, $q(z)$, is constant in $x,y$ and has
boundary conditions $q(-\infty)=0$ and $q(+\infty)=q_0$. The
profile $q(z)$ has constant phase since the potential is a real
function that depends only on $|q|$. In the Higgs phase there is a
vortex obtained by choosing an element $n$ of the homotopy group
$[n]\in \pi_1(\mathbf{S}^1)$: \bea \label{vortex}  q &=& e^{i n
\theta} q(r) \ , \\  A_{\theta} &=& \frac{n}{e r} A(r) \ ,
\nonumber \eea where we have used cylindrical coordinates
$r,\theta,z$ instead of $x,y,z$. Clearly the wall and the tube are
continuously related. Consider a kink in the radial direction at a
radius much greater than its thickness. We can roll it around the
axe $z$ giving it a gauge rotation $e^{in\theta}$. This
automatically turns on a pure gauge field at large distance that,
for continuity, must have a flux inside. This configuration in
general is not a minimum of the energy density, so it will start
to loose energy until it reach the stable vortex.

\subsubsection{The wall limit \label{wallim}}

The differential equations for the profile functions of the vortex
(\ref{vortex}) are \bea \label{eqdiff} && \frac{\d^2 q}{\d
r^2}+\frac{1}{r}\frac{\d q}{\d r}-n^2
\frac{(1-A)^2}{r^2}q-\frac{\delta V}{\delta q}=0 \ , \\ &&
\frac{\d^2 A}{\d r^2}-\frac{1}{r}\frac{\d A}{\d r}+e^2(1-A)q^2=0 \
, \nonumber \eea where $n$ is the winding number. We are looking
for some limit of the parameters so that the vortex really looks
like a rolled wall. In this limit the profile functions should be:
\bea  \label{theta} q(r) &=& q_0 \theta_H(r-R_V)\ ,  \\
A(r)&=&r^2/R_V^2  \qquad  0 \leq r \leq R_V \ , \nonumber \\
A(r)&=&1\  \qquad  r \geq R_V \ , \nonumber \eea  where $\theta_H$
is the step function. First of all we manipulate a bit the
differential equations (\ref{eqdiff}) to simplify them.  The
potential can be written as a dimensionless function \beq V(q)=v_0
\V \left(\frac{q}{q_0}\right)\ , \eeq where $v_0$ is the scale of
the potential and $q_0$ the vev. In the following we will not use
the shape of the dimensionless potential $\V$. The only important
thing is that $\V(0)=\V(1)=0$ and its height is of order $1$ (the
simplest example is $\V(\chi)=\chi^2(\chi^2-1)^2$). We also scale
the scalar field $q=q_0 \chi$. After these scalings the equations
(\ref{eqdiff}) for the profiles becomes: \bea \label{eqscal1} &&
\frac{\d^2 \chi}{\d r^2}+\frac{1}{r}\frac{\d \chi}{\d r}-n^2
\frac{(1-A)^2}{r^2} \chi- \alpha \frac{\delta \V}{\delta \chi}=0 \
, \\  \label{eqscal2} && \frac{\d^2 A}{\d r^2}-\frac{1}{r}\frac{\d
A}{\d r}+ \beta  (1-A)\chi^2=0 \ . \eea So there are three
parameters that enter the game: \beq n\ , \qquad
\alpha=\frac{v_0}{{q_0}^2}\ , \qquad \beta=e^2 {q_0}^2\ .\eeq The
domain wall  tension and  thickness are respectively:\footnote{For
this qualitative result it is sufficient to take the wall
lagrangian ${\cal L}=-\de q \de q-V(q)$ and then write the tension
as a function of the thickness $ T(\Delta) \sim
\frac{q_0}{\Delta}+v_0 \Delta$. Minimizing $T(\Delta)$ with
respect to $\Delta$ we find (\ref{qualitative}).}
 \beq \label{qualitative} T_W \sim  q_0 \sqrt{v_0}\ , \qquad \Delta_W
\sim  \frac{q_0}{\sqrt{v_0}} \ .\eeq

Now comes the first non trivial hint. If the wall limit exists,
then formula (\ref{eqscal}) can be trusted in this limit. But the
radius of the vortex $R_V$ comes out from equations (\ref{eqdiff})
and so must depend only on the three relevant parameters
$n,\alpha,\beta$. In general a function of $n,e,v_0,q_0$ cannot be
expressed as a function of $n,\alpha,\beta$, but for
(\ref{eqscal}) this is possible: \beq R_V \sim \frac{n^{2/3}}
{\alpha^{1/6} \beta^{1/3}} \ . \eeq If we wouldn't have found such
expression, we would have concluded that the wall limit doesn't
exist. This result encourages us to go on.

\begin{figure}[h!t]
\begin{center}
\leavevmode \epsfxsize 14 cm \epsffile{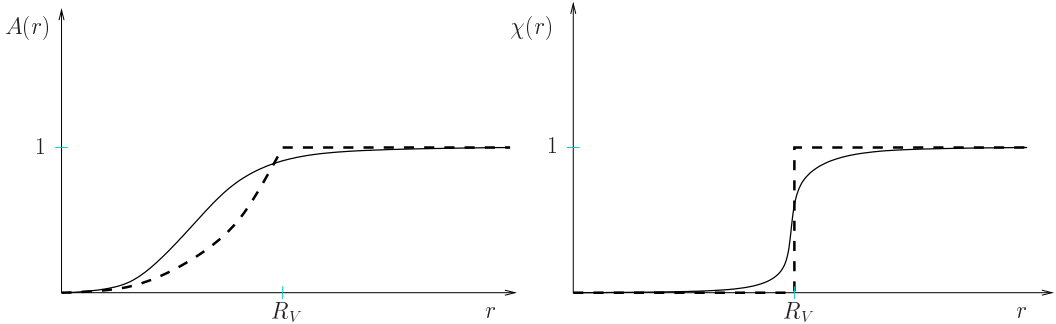}
\end{center}
\caption{\footnotesize Increasing the winding number $n$ and
keeping fixed the radius of the vortex $R_V$, the profile
functions approach the wall vortex.} \label{limite}
\end{figure}

Now we discuss the wall limit. In this paragraph we look for a
limit in which the radius of the vortex remain constant and the
solution approaches the wall vortex (\ref{theta}) (our goal is
described in Figure \ref{limite}). 
The $\chi$ profile must become a step function: it is zero inside
$R_V$, it goes from zero to one in a distance $\Delta_W$, and then
remains constant at one.  Thus $\chi''$ in equation
(\ref{eqscal1}) develops a $\delta'(r-R_V)$ singularity or,
thinking in terms of $\Delta_W$, $\chi'' \sim 1/{\Delta_W}^2$.
Thus to counterbalance this divergence in (\ref{eqscal1}) we must
have $\alpha$ that goes to infinity like $1/{{\Delta_W}^2}$.  The
encouraging fact is that $\alpha \frac{\delta \V}{\delta \chi}$
really resembles a $\delta'(r-R_V)$ in this limit (note that this
is true only if the two vacua are exactly degenerate).  Now
consider the second equation (\ref{eqscal2}) where $A''(R_V)$ has
a $(2 / R_V) \delta(r-R_v) $ singularity, or thinking in terms of
$\Delta_W$, $A''(R_V) \sim 1/(R_V \Delta_W)$. Since $(1-A)\chi$ is
of order $\Delta_W / R_v$ around $R_v$, we must also send $\beta$
to infinity like $1/{\Delta_W}^2$.  Now we can make the conjecture
of the wall limit.
\newline {\bf Mathematical Conjecture:}  { \it Consider the
succession of parameters $\alpha_n=n^{4/3} \alpha_1$ and $
\beta_n=n^{4/3} \beta_1$ and call the solution of (\ref{eqscal1})
and (\ref{eqscal2}) with the vortex boundary conditions,
$\chi_{n,\alpha_n,\beta_n}(r)$ and $A_{n,\alpha_n,\beta_n}(r)$. In
the limit $n \to \infty$ \bea && \lim_{n \to \infty}
\chi_{n,\alpha_n,\beta_n}(r) \to \theta_H(r-R_v)\ , \\ && \lim_{n
\to \infty} A_{n,\alpha_n,\beta_n}(r) \to \left\{
\begin{array}{cc} r^2/{R_V}^2 & 0 \leq r \leq R_V \ , \\
1 & r>R_V  \ . \\ \end{array}\right. \nonumber \eea } \newline
This limit has been chosen so that the radius of the vortex remain
constant $R_V \sim n^{2/3} {\alpha_n}^{-1/3}
{\beta_n}^{-1/6}={\alpha_1}^{-1/3} {\beta_1}^{-1/6}$ and also the
ratio $\alpha_n / \beta_n$ remains constant. The ratio $\alpha /
\beta$ disappears in the wall limit and is related only to the
shape of the limiting functions $\chi_{n,\alpha_n,\beta_n}(r)$ and
$A_{n,\alpha_n,\beta_n}(r)$. Probably a stronger version of the
conjecture is  true:  the ratio $\alpha_n /\beta_n$ is kept
limited from above and from below during the limit, so that it
doesn't go neither to infinity nor to zero.

The limit discussed above has been called mathematical since it is
easy to express in a mathematical language. In this paragraph we
discuss a more physical situation in which we keep fixed the
parameters of the theory and we only change the winding number
$n$. When $n$ is increased the radius of the vortex becomes large
while the thickness of the wall remains constant since it doesn't
depend on $n$. We are going to use (\ref{eqscal}) in an self
consistent way. Suppose that $n$ is enough large so that
(\ref{eqscal}) is true, then we use the expression of $R_V$ to
obtain the condition for $n$ so that $R_V \gg \Delta_W$.
\newline {\bf Physical Conjecture:} {\it We keep $\alpha$ and $\beta$ fixed
and increase $n$. The condition $R_V \gg \Delta_W$ is $n \gg
(\beta/\alpha)^{1/4}$. When this is satisfied the vortex resembles
a wall vortex and its tension is given by \beq \label{spectrum}
T_V=3  \sqrt[3]{2} \pi \; \left(\frac{n {T_W}}{e}\right)^{2/3} \ .
\eeq}

\subsubsection{Pushing a monopole against the wall}

\begin{figure}[h!t]
\begin{center}
\leavevmode \epsfxsize 11 cm \epsffile{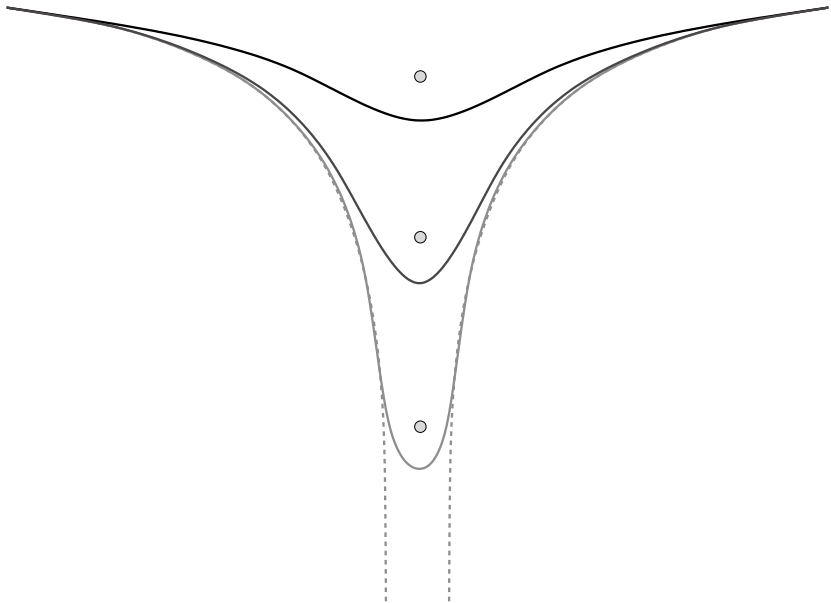}
\end{center}
\caption{\footnotesize A monopole in the Coulomb phase is pushed
against that domain wall. At the end of the process we are left
with a monopole in the Higgs phase confined on a vortex that ends
on the wall.} \label{push}
\end{figure}

There is another intuitively way in which we can see the wall
continuously transformed into the vortex. Consider a monopole in
the Coulomb phase. The lines of its magnetic flux will be tangent
to the wall surface. Now we move the monopole in the wall
direction as in Figure \ref{push}. 
The wall is repelled and change its shape. Far from the $x,y$
coordinate of the monopole the shape of the wall is logarithmic $z
\propto \log{(x^2+y^2)}$. If we continue to push the monopole in
the negative $z$ the configuration will resemble more and more a
monopole in Higgs phase, attached to a flux tube that ends on the
wall. This physical picture well explains a wall that is
continuously transformed into a vortex. It's also clear that this
kind of vortices can naturally end on the wall that has created
them since they are made of the same stuff.

A more detailed and quantitative analysis of this junction will
appear in \cite{ioeken}.

\subsubsection{$\N=2$ super QED broken to $\N=1$ \label{susy}}

Now we consider $\N =2$ SQED broken to $\N=1$ by a superpotential
for the adjoint field.  The $U(1)$ gauge multiplied is composed by
the superfields $W_{\alpha}$ and $\Phi$, while the matter
superfields are $Q$ of charge $+1$ and $\tQ $ of charge $-1$. The
Lagrangian is the following: \bea \label{classicalSQED}
{\cal L}&=&\int d^2\theta \, \frac{1}{4e^2}W^{\a}W_{\a}+h.c.\\
&&+\int d^2\theta d^2\bar{\theta} \, (\frac{1}{e^2}\Phi\dag\Phi+Q\dag e^{V}Q+{\widetilde Q}\dag e^{-V}{\widetilde Q})\nonumber\\
&&+\int d^2\theta \, \sqrt{2} ({\widetilde Q}\Phi Q-m{\widetilde
Q}Q+W(\Phi))+h.c.\ .\nonumber \eea The potential for the scalar
fields is \beq \label{pot}
V=2|(\phi-m)q|^2+2|(\phi-m)\tq|^2+2e^2|\tq
q+W\p(\phi)|^2+\frac{e^2}{2}(|q|^2-|\tq|^2)^2\ . \eeq This
potential has two kind of minima: one is the Higgs vacuum \beq
\phi=m \ , \qquad |q|=|\tq| \ ,  \qquad \tq q = - W\p(m)\ , \eeq
and the others are a set of Coulomb vacua, each for every
stationary point of the superpotential \beq \phi=a_i\ ,\qquad
q=\tq=0 \ . \eeq

In the Higgs vacuum the gauge group $U(1)$ is completely broken by
the quark condensate, so the theory admits  vortex solutions that
belongs to the homotopy group $\pi_1(U(1))=\mathbb{Z}$. We don't
loose any information choosing \beq \tq = -q\dag
\frac{W\p(\phi)}{|W\p(\phi)|} \ . \eeq We are left with the
following theory: \beq {\cal
L}=-\frac{1}{4e^2}F_{\mu\nu}F^{\mu\nu}-\frac{1}{e^2}
\de_{\mu}\phi^{\dagger}\de^{\mu}\phi-(D_{\mu}q)^{\dagger}(D^{\mu}q)-V(\phi,|q|)\
, \eeq where the potential is \beq
V(\phi,|q|)=2|(\phi-m)q|^2+\frac{e^2}{2} (|q|^2-2|W\p(\phi)|)^2  \
.\eeq The vortex solution in terms of the profile functions is:
\bea   q &=& e^{i n \theta} q(r) \ , \\
\phi &=& \phi(r) \ , \nonumber \\  A_{\theta} &=& \frac{n}{e r}
A(r) \ . \nonumber \eea The boundary conditions for $q(r)$ and
$A(r)$ are the usual ones. The boundary conditions for $\phi(r)$
are $\phi(\infty)=m$ and $\phi(0)=a_j$ where the root $a_j$ will
be the one that makes the vortex lighter.  When all the roots
$a_i$ are reals, it is clear that $a_j$ is the one that minimize
$|m-a_i|$. This vortex can be considered as made off the wall
connecting the Higgs vacuum and the $j$-Coulomb vacuum. It is thus
straightforward to generalize the physical conjecture made in
\ref{wallim}. For $n$ sufficiently large so that $R_V \gg
\Delta_W$, the spectrum of these vortices is (\ref{spectrum}).

{\it Almost-BPS solution}

The theory under consideration has been widely studied since it
arise as an effective description of more complicated nonabelian
theories \cite{SW,DS,vainyung,Hanany:1997hr}. This SQED is
sometimes magnetic dual with respect to the original degrees of
freedom and thus our vortices describes confinement of quarks.
Usually these vortices have been studied in an almost BPS regime,
that is by neglecting the second derivative of the superpotential.
In this limit we can make a stronger ansatz for the vortex
solution:
 \beq \label{BPSansatz}
\phi=m \ . \eeq  Thus the Lagrangian becomes the usual one in the
BPS limit \cite{Bogomolny:1975de,Edelstein:1993bb}: \beq
{\cal
L}=-\frac{1}{4e^2}F_{\mu\nu}F^{\mu\nu}-(D_{\mu}q)^{\dagger}(D^{\mu}q)-\frac{e^2}{2}(|q|^2-2|W\p|)^2
\ . \eeq The spectrum of vortices is the well known BPS
proportionality between tension and charge: \beq
\label{BPSsemiclassical} T=4\pi| n W\p(m) |\ . \eeq

\begin{figure}[h!t]
\begin{center}
\leavevmode \epsfxsize 10 cm \epsffile{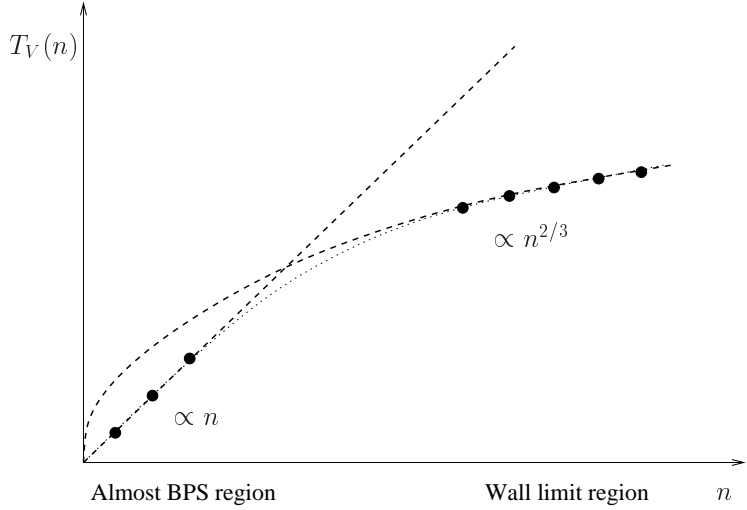}
\end{center}
\caption{\footnotesize Spectrum of vortices in $\N=2$ SQED broken
by a superpotential. If $\frac{e^2{{W\p}\p}^2}{W\p} \ll 1$ an
almost BPS region is present for small winding numbers. Increasing
$n$ non-BPS corrections become larger and not negligible. When the
winding $n$ is sufficiently big we are in the wall vortex regime.}
\label{winding}
\end{figure}

Clearly (\ref{BPSansatz}) is only an approximation. It is enough
to look at the equation of motion for $\phi$ and see that it is
not satisfied inside the vortex: \beq \frac{\Box
\phi}{e^2}=2(\phi-m)|q|^2+e^2 {W\p}\p(\phi)\dag
(|q|^2-2|W\p(\phi)|)\frac{W\p(\phi)}{|W\p(\phi)|}\ . \eeq An
important point is to find the condition under which
(\ref{BPSsemiclassical}) is a good approximation to the spectrum.
In \cite{Bolognesi:2004yh} we have done this for the winding
number $n=1$, and the condition has been found to be: \beq
\label{goodone} \frac{e^2{{W\p}\p}^2}{{W\p}} \ll 1\ . \eeq It is
clear also from \cite{Bolognesi:2004yh} that as the winding $n$
becomes bigger, so do the non-BPS corrections. For consistency
with the wall vortex limit, we should have a spectrum like Figure
\ref{winding}.
For sufficiently small $n$ the tension is given by the BPS formula
(\ref{BPSsemiclassical}) and is proportional to $n$. For
sufficiently big $n$ the tension is given by (\ref{spectrum}) and
is proportional to $n^{2/3}$. There must not be superposition
between the two regions.

{\it $\N=2$ SQED broken by adjoint mass}

 Now we make the consistency check the the two regions do not
overlapped. For simplicity we limit ourselves to the quadratic
superpotential $ W(\Phi)=\frac{1}{4} \mu \Phi^2 $. The energy
density for the wall is
 \beq
T=\int \d x  ( \frac{1}{e^2}\de_x \phi\dag \de_x \phi+ \de_x q\dag
\de_x q+ 2|(\phi-m)q|^2+ \frac{e^2}{2}(|q|^2 -\mu \phi)^2 ) \ .
\eeq We rewrite it in terms of the dimensionless variables $
\phi=m\varphi$ and $q=\sqrt{\mu m}\chi$:
 \beq \label{tt}
T=\int \d x  ( \frac{m^2}{e^2}\de_x \varphi\dag \de_x \varphi+
\de_x \chi\dag \de_x \chi+ 2m^3\mu|(\varphi-1)\chi|^2+ \frac{e^2
m^2 \mu^2}{2}(|\chi|^2 -\varphi)^2 ) \ . \eeq The condition for
the existence of an almost BPS region is
 \beq \label{alm} \frac{e^2 \mu}{m} \ll 1 .\eeq
 The $n=1$ vortex would also belong to the wall vortex region is the condition $R_V \gg \Delta_W$ would be satisfied.
 In terms of our variables this condition would be
 \beq  \label{wal} \frac{1}{e^{2/3}{T_W}^{1/3}} \gg \Delta_W .\eeq
 We are going to prove that, if (\ref{alm}) is satisfied, then (\ref{wal}) cannot be
 satisfied.

Under the assumption that (\ref{alm}) is satisfied, the tension
(\ref{tt}) can be approximated by \beq \label{var} T \sim
\frac{m^2}{e^2\Delta}  +m^3\mu \Delta \ , \eeq where $\Delta$ is
the length scale of variation of the fields. Minimizing
(\ref{var}) with respect to $\Delta$  we obtain: \beq \Delta_W
\sim \frac{1}{e\sqrt{\mu m}}\ ,\qquad T_W \sim
\frac{m^{5/2}\mu^{1/2}}{e} \ .\eeq Thus the condition for the wall
vortex is \beq \frac{1}{e^{2/3} {T_W}^{1/3} \Delta_W} \sim
\left(\frac{e^2 \mu}{m}\right)^{1/3} \gg 1 \ ,\eeq and it is
clearly not satisfied if (\ref{alm}) is true.

\subsubsection{Quantum corrections}

The non-supersymmetric theory (\ref{BasicVortex}) with the
potential of Figure \ref{potential} has a problem: it is not
stable under quantum corrections since there is no spontaneously
broken discrete symmetry that can relate an Higgs vacuum with a
Coulomb vacuum. Only with a fine tuning of the parameters we can
have this degeneracy.

Degeneracy of vacua, not related by any spontaneously broken
discrete symmetry, is a common feature of supersymmetric gauge
theories. In fact in the supersymmetric theory described in
\ref{susy} the degeneracy is protected by non renormalization
theorems. We can thus ask what happens to the relations
(\ref{eqscal}) when we consider quantum corrections. Note that to
obtain (\ref{eqscal}) we have used only general principles and so
they should be valid also in a full quantum theory, even if the
various parameters that enter in the game $T_V, R_V, T_W,
\Delta_W$ are subject to quantum corrections. Special attention
should be given to the coupling constant $e(\mu)$ since in the
full quantum theory depends on the energy scale $\mu$. But since
we want to study vortices with large $n$ and so large radius
$R_V$, we should keep the low-energy coupling constant
$e(m_{lightest})$ where $m_{lightest}$ is the mass of the lightest
charged particle.

\subsubsection{Zero modes}

\begin{figure}[h!t]
\begin{center}
\leavevmode \epsfxsize 10 cm \epsffile{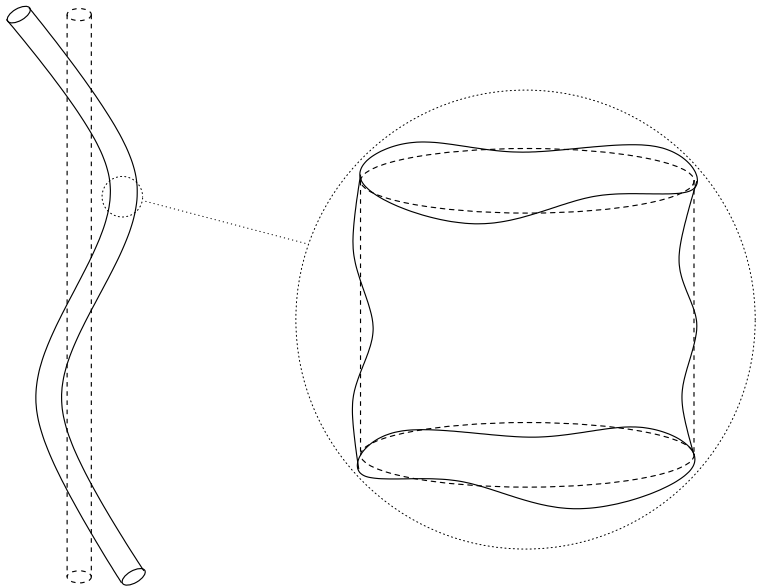}
\end{center}
\caption{\footnotesize Fluctuation of the displacement zero modes.
For wave length $\lambda \gg R_V$ we have a vibrating string. For
wave length $\Delta_W \ll \lambda \ll R_V$ we have a vibrating
wall compactified on a circle.} \label{zeromodes}
\end{figure}

The low-energy dynamics of a $p$-soliton is described by an
effective $p+1$ dimensional field theory on the world volume of
the soliton. By low-energy we mean length greater that the
thickness of the soliton. The target space of the effective action
is the space of zero modes. this space can be divided in
displacement zero modes, the ones that describe fluctuation of the
brane in the space, and internal zero modes. Up to now it is the
standard technique for describe soliton dynamics. In the case of
the wall vortex it is not hard to imagine that there should be a
relation between the effective $2+1$ theory of the domain wall and
the effective $1+1$ theory of the vortex. In fact the vortex
dynamics is described by the $2+1$ theory of the wall with a
spatial direction compactified on a circle of radius $R_V$. For
wave lengths $\lambda$ so that $\Delta_W \ll \lambda \ll R_V$,  is
a vibrating membrane and for wave lengths $\lambda \gg R_V$ it is
a vibrating string. In Figure \ref{zeromodes} we have the
displacement zero modes in the two different regimes. It is also
clear that the wall vortex share the same internal zero modes
target space with the wall that makes it.

\subsection{Nonabelian theories and the existence at strong
coupling \label{naexistence}}

It is a natural question if these wall vortices can arise in
nonabelian theory and at strong coupling. The answer is yes and,
as we will see in a moment, it is enough to consider confining
vacua with different confinement index.

Take a generic $SU(N)$ gauge theory.  The confinement index $t$ is
defined as the minimal integer so that $Q^t$ in unconfined. By
$Q^t$ we mean the representation of $t$ antisymmetrized quarks $Q$
in the fundamental of the gauge group. When  $t=N$ the theory is
completely confined, like ordinary QCD. When $t=1$ the theory is
completely unconfined. $t$ must be a divisor of $N$. The
confinement is caused by strings with topological number
$\mathbb{Z}_t$. A $k$-string, with $k \in \mathbb{Z}_t$, confines
the $Q^k$ representation. The $t$-string is trivial and in fact
$Q^t$ is unconfined.

 Now consider a case in which this theory has
one vacuum $A$ with confinement index $t_A$ and another vacuum $B$
with confinement index $t_B > t_A$. Suppose that exists a domain
wall interpolating between these two vacua. There will be at least
one representation $Q^k$ that is confined in one vacuum and not
confined in the other. For example $Q^{t_A}$ is not confined in
$A$ but confined in $B$.   We can make a continuous
transformation, such as the one previously described in Figure
\ref{push}. We take a $Q^{t_A}$ particle in vacuum $A$ and we push
it against the wall. We will end with a $Q^{t_A}$ in vacuum $B$
connected to a $t_A$-string that ends on the wall. This continuous
interpolation between the wall and the string shows that the
$t_A$-string is indeed a wall vortex.

In general every $k$-string in vacuum $B$ is a wall vortex with
respect to the $A$/$B$ wall if $k$ is a multiple of $t_A$ but not
of $t_B$. The same can be said on the other side of the wall.
Every $k$-string in vacuum $A$ is a wall vortex with respect to
the $B$/$A$ wall if $k$ is a multiple of $t_B$ but not of $t_A$.
To be more precise we should think of the strings as living in an
extended group $\mathbb{Z}_{\lcm{(t_A,\,t_B)}}$ that contains both
$\mathbb{Z}_{t_A}$ and $\mathbb{Z}_{t_B}$. Strings in vacuum $A$
that belong to the subgroup $\mathbb{Z}_{\lcm{(t_A,\,t_B)}/t_B}$
are wall vortices with respect to the $A$/$B$ wall. Strings in
vacuum $B$ that belong to the subgroup
$\mathbb{Z}_{\lcm{(t_A,\,t_B)}/t_A}$ are wall vortices with
respect to the $B$/$A$ wall. We will came back to this point in
 \ref{super}.

\begin{figure}[h!t]
\begin{center}
\leavevmode \epsfxsize 9 cm \epsffile{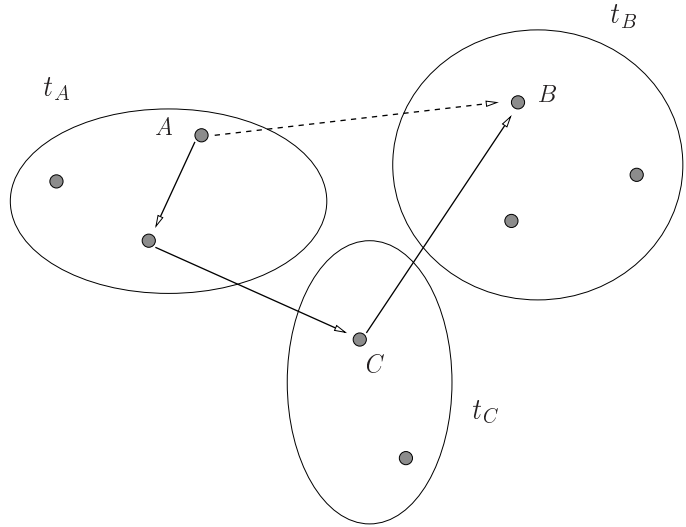}
\end{center}
\caption{\footnotesize In general it not true that given any
couple of vacua $A$ and $B$ a domain wall that interpolates
between them exist. But if it doesn't exist then it decays in a
path of stable domain walls that connect $A$ and $B$ passing
through other vacua. If the $t_A \neq t_B$, surely there will be a
stable domain wall at some point of the path that connects vacua
with different confinement index. } \label{wall}
\end{figure}

In general we can prove that a domain wall interpolating between
vacua with different confinement index exists. Consider the set of
vacua of the theory and group them in sets with the same
confinement index like in Figure \ref{wall}.
It is not true that for any couple of vacua $A$ and $B$ the domain
wall interpolating between them exist. But, for sure, it will
exist at least a path connecting the two vacua so that in every
segment of the path the domain wall exist and it's stable. In this
path there will be some domain wall interpolating between vacua
with different confinement index.

The study of the wall limit for nonabelian strings, such as the
one done in \ref{wallim}, will appear in \cite{miononabeliani}.

\section{Classification of the Domain Wall/Flux Tube Junctions \label{classification}}

\begin{figure}[h!t]
\begin{center}
\leavevmode \epsfxsize 10 cm \epsffile{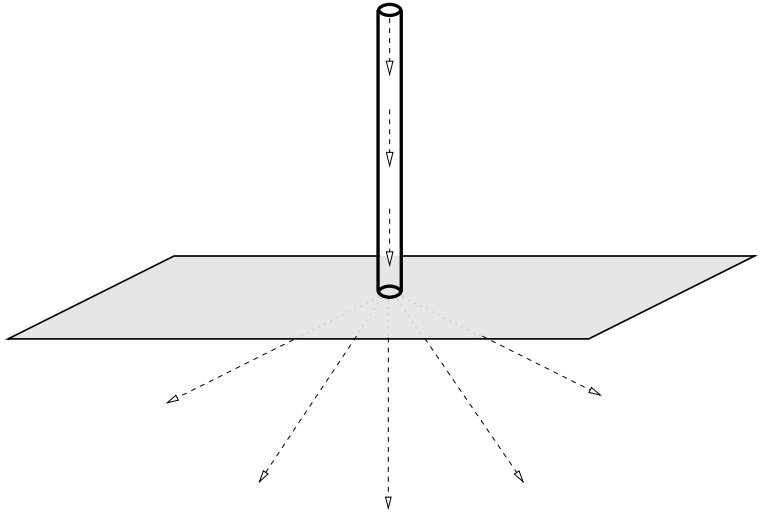}
\end{center}
\caption{\footnotesize The Coulomb junction. A vortex ends on the
wall and the flux is spread into the other half space. The point
where it ends is seen in the Coulomb vacuum as a monopole of
double charge.} \label{Coulomb}
\end{figure}

The wall vortex studied in  Section \ref{made} is one particular
case of wall/tube junction. The purpose of this Section is to
classify the various kind of junctions between walls and tubes.
The junctions can be distinguished by simple properties:
\begin{center}
\begin{itemize}
\item If the tube can end or not on the wall. \item Where the flux
goes into.
\end{itemize}
\end{center}
In this spirit, the Coulomb junction of Section \ref{made}  can be
schematically represented as in Figure \ref{Coulomb}. The tube
ends on the wall and the flux goes in the opposite half space.
 To analyze
other examples we make the first distinction. In \ref{local} we
consider vortices that can end on the wall. In \ref{crossing} we
consider vortices that cross the wall and continue on the other
side.

\subsection{The Flux Tube Ends on the Wall \label{local}}

In string theory, a string can have Dirichlet boundary conditions
if it ends on a dynamical object called D-brane. By analogy, in
some QFT, similar phenomenon can happen when a flux tube ends on
the wall.  We will briefly describe two examples in which a flux
tube can terminate on a wall. In the first case there is a
localization of the flux on the wall. Thus the tube spread out its
flux into the wall.  In the second example there is a bound state
confined on the wall that captures the flux. The bound state is an
object composed by two pieces that belong to the two different
vacua.

\subsubsection{\label{nahiggs} Localization of the flux}

\begin{figure}[h!t]
\begin{center}
\leavevmode \epsfxsize 10 cm \epsffile{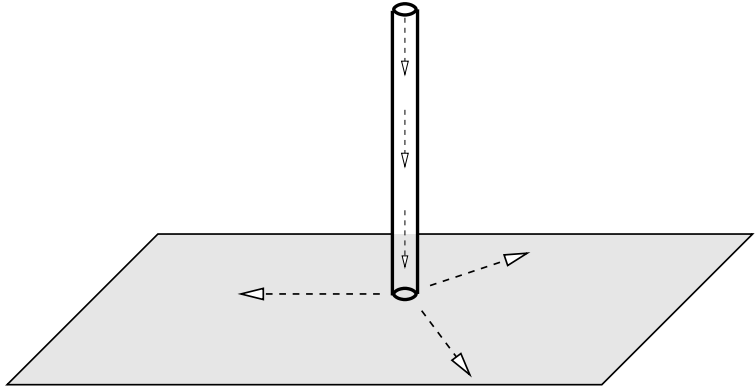}
\end{center}
\caption{\footnotesize The D-brane junction. A vortex ends an a
wall and spread its flux into it. The effective theory on the wall
must contain massless gauge field. The point where the tube ends
is an electric charge with respect to the $2+1$ effective gauge
theory.} \label{Local}
\end{figure}

This case is probably the more studied, particularly in the last
few years. We have a domain wall interpolating between two vacua
both in the Higgs phase. The magnetic strings on both side of the
wall can end on it and spread the flux inside it (see Figure
\ref{Local}).
Inside the wall there must be a localized gauge theory. Note that
if we are able to prove that the $2+1$ dimensional theory contains
the massless gauge field, the confining string must end on the
wall (the other possible configurations would be not favorite
energetically).

Recently there has been research in the the so called non-abelian
Higgs model \cite{Shifman:2002jm,nahiggs}. It is a $U(N_c)$ $\N=2$
gauge theory with $N_f$ hypermultliplets of mass $m_i$. The theory
is then broken to $\N=1$ adding a Fayet-Iliopoulos term for the
$U(1)$ subgroup. When $N_f \geq N_c$ the theory has
$\binom{N_f}{N_c}$ vacua since every color must be locked to some
flavor. This theory is a nice lab because it possesses all the
three kinds of solitons: monopoles, vortices and domain walls,
that can appear in various junctions that preserves supersymmetry.
Consider in particular a fundamental wall that separates two vacua
$\dots,i,\widehat{i+1},\dots$ and $\dots,\widehat{i},i+1,\dots$
(with the hat we indicate the flavors that are not locked). In the
first vacua a $U(1)$ is locked to the flavor $i$ while the flavor
$i+1$ is unlocked, the opposite in the other vacua. In both vacua
the $U(1)$ is Higgsed, in the first by the condensation of the
flavor $Q_i$ and in the second by the flavor $Q_{i+1}$. They admit
a magnetic vortex that breaks $1/2$ of supersymmetries and
saturates the BPS bound. The central charge for the vortex is
given essentially by the Fayet-Iliopoulos term. It has been shown
that the walls are also $1/2$-BPS and in particular the one under
consideration (from $i, \widehat{i+1}$ to $\widehat{i}, i+1$), has
a Coulomb phase inside with respect to this $U(1)$. The junction
vortex-wall has been shown to be $1/4$-BPS. For generalizations to
nonabelian strings and nonabelian walls see \cite{nonabelian}.

\subsubsection{\label{sy} Bound state localized on the wall}

\begin{figure}[h!t]
\begin{center}
\leavevmode \epsfxsize 13 cm \epsffile{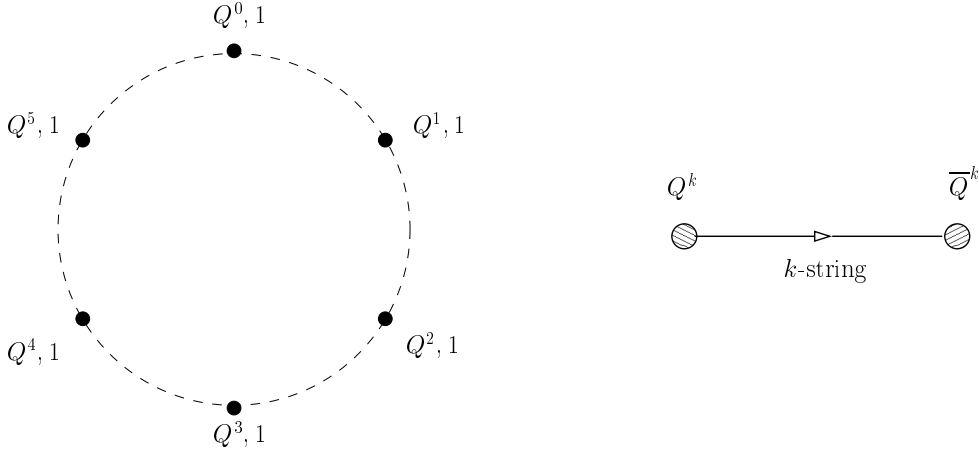}
\end{center}
\caption{\footnotesize The $\N=1$ pure Super Yang Mills has $N$
different vacua labelled with $r$. The confinement in each vacuum
is explained by the condensation of a particle that carries
magnetic charge $1$ and electric charge $Q^r$. In every vacuum
strings belong to the topological group $\mathbb{Z}_N$. A
$k$-string connect a $Q^k$ with a $\overline{Q}^k$ and creates a
meson.} \label{SYM}
\end{figure}

Another example of strings anding on walls is provided by $\N=1$
$SU(N)$ super Yang-Mills. This theory has $N$ vacua obtained by
the spontaneous breaking of the anomaly free residual
$R$-symmetry. We label these vacua with the index $0 \leq r < N$
(see Figure \ref{SYM}).
Every vacuum has confinement index $t=N$ and so there are
$\mathbb{Z}_{N}$ strings responsible for the confinement of
sources in any non trivial representation of the center of the
gauge group. The string $k \in \mathbb{Z}_{N}$ confines the
representation $Q^k$.   The confinement in the $r$ vacua can be
understood as caused by the condensation a monopole bounded
together with a $Q^r$. We indicate this object with $(Q^r,1)$
where the $1$ refers to the magnetic charge.

\begin{figure}[h!t]
\begin{center}
\leavevmode \epsfxsize 8 cm \epsffile{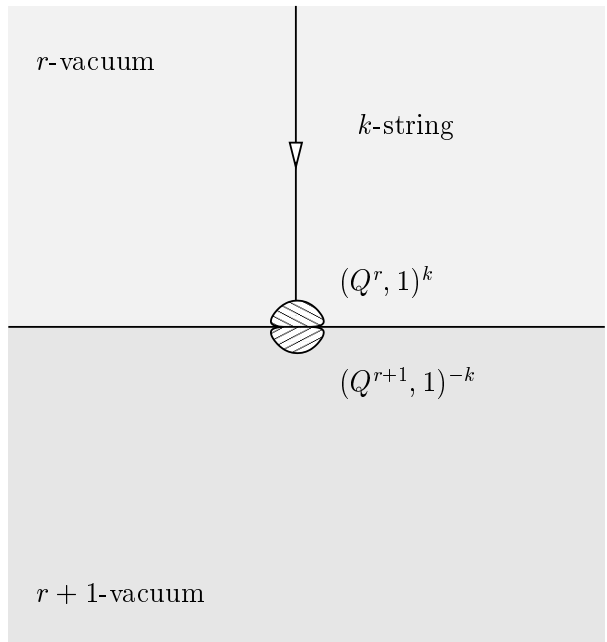}
\end{center}
\caption{\footnotesize The End junction. The vortex ends on a
boundstate confined on the wall. } \label{boundSYM}
\end{figure}

It has been shown in \cite{Witten:1997ep} that the domain
wall\footnote{ Domain walls in $\N=1$ super Yang-Mills have been
studied in \cite{Dvali:1996xe,Shifman,Dvali}. The effective action
on these walls has been found in \cite{Acharya:2001dz}.}
interpolating between two adjacent vacua $r$ and $r+1$ is a
D-brane for the $k$-strings. This phenomenon happens also for
every wall interpolating between a vacuum $r_1$ and a vacuum $r_2$
so that $r_2-r_1$ is prime with $N$. The easiest way to understand
it is the Rey's argument reported in \cite{Witten:1997ep}. The
confinement in the $r$-vacuum is due to the condensation of a
$(Q^r,1)$ particle. To prove that the $k$-string can end on the
wall we need a bound state confined on the wall that carries the
same charge of $Q^{-k}$. As showed in Figure \ref{boundSYM} this
object can be obtained putting together $k$ $(Q^r,1)$ and $-k$
$(Q^{r+1},1)$ on the opposite sides of the
wall.\footnote{Actually, to have a rigorous prove, we should show
the composite object has a binding energy.}

\subsection{The Flux Tube Crosses the Wall \label{crossubsection}}

\begin{figure}[h!t]
\begin{center}
\leavevmode \epsfxsize 9 cm \epsffile{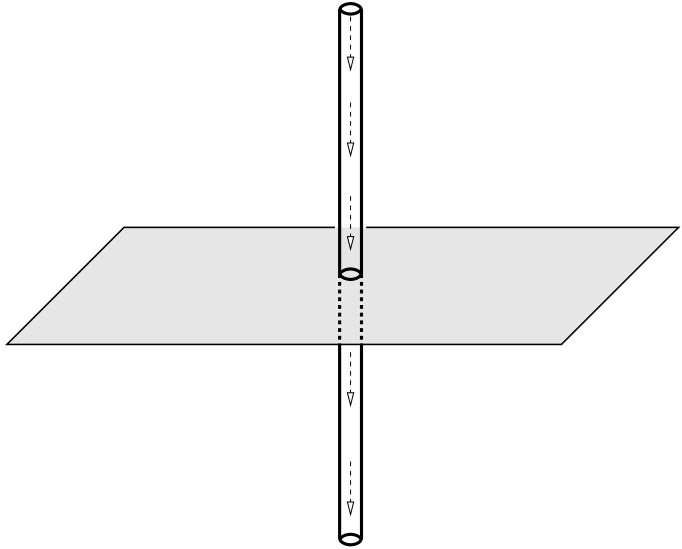}
\end{center}
\caption{\footnotesize The Cross junction. The flux tube crosses
the wall and becomes a vortex in the other vacuum.}
\label{crossing}
\end{figure}

When a  flux tube ends on a wall, the flux must go somewhere.
There are only two possibility: it goes on the other half space
(see Section \ref{made}), or it is confined inside the wall (see
 \ref{local}). Now we consider the case in which the
vortex cannot end on the wall and so  must  continue into another
vortex in the other vacua (see Figure \ref{crossing}).
They could also bring different fluxes since there could be a
particle localized on the wall, in the junction between the two
vortices, that carries the difference of the fluxes.

The first example is the trivial one. Take the usual $U(1)$ gauge
theory coupled to a charged scalar in the Higgs phase. Than add a
real scalar field responsible for the creation of a domain wall.
We don't put interactions between the sector of the theory
responsible for the vortex and the sector responsible for the
wall. The domain wall is transparent to the vortex that can cross
it without any modification. This trivial junction also arise in
the non-abelian Higgs model considered in \ref{nahiggs}.  For
example, we saw that the wall interpolating between the  vacua
$\dots,i,\widehat{i+1},\dots$ and $\dots,\widehat{i},i+1,\dots$,
forms a D.brane junction with respect to the vortex that carry the
flux of the $U(1)$ that changes locking. The same wall makes a
trivial junction with respect to the other $U(1)$'s that are left
unchanged.

\subsubsection{A vortex with a wall around it}

\begin{figure}[h!t]
\begin{center}
\leavevmode \epsfxsize 9 cm \epsffile{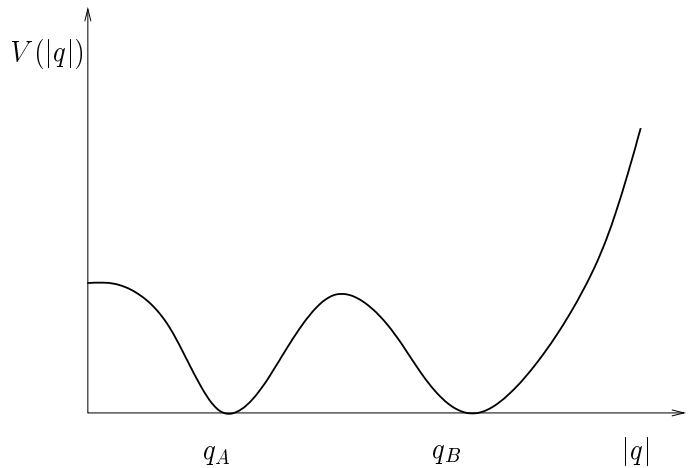}
\end{center}
\caption{\footnotesize A potential with two degenerate Higgs
vacua.} \label{wallaroundpot}
\end{figure}

\begin{figure}[h!t]
\begin{center}
\leavevmode \epsfxsize 9 cm \epsffile{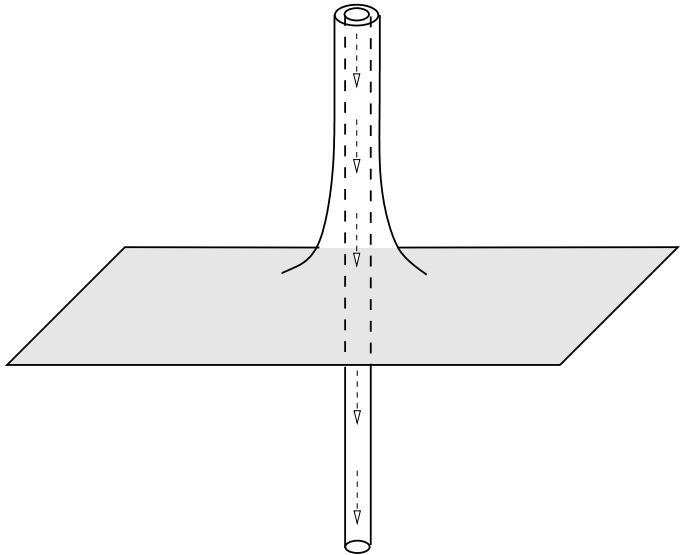}
\end{center}
\caption{\footnotesize A vortex crosses the wall and becomes a
heavier vortex that can be though as the previous one with a wall
wrapped around.} \label{wallaroundfig}
\end{figure}

Now we move to a non trivial case. Consider the $U(1)$ gauge
theory (\ref{BasicVortex}), where the potential has two different
 minima at $q_A < q_B$ (see Figure \ref{wallaroundpot}).
There is a wall that interpolates between the two vacua, that has
a profile $q=q(z)$ with boundary conditions $q(-\infty)=q_A$ and
$q(+\infty)=B$. The wall has not have a Coulomb phase inside it,
since $q(z)$ is always different from zero.  Both the vacua $A$
and $B$, admit magnetic flux tubes and the tension $T_A$ will be
smaller than the tension $T_B$.  The Coulomb junction is excluded
since $A$ and $B$ are both in the Higgs phase. The D-brane
junction is also excluded since there is not a Coulomb phase
inside the wall. The only possibility is that the vortex in $A$
crosses the wall and becomes the vortex in $B$ (see Figure
\ref{wallaroundfig}).
Since they carry the same flux there non need of any particle
localized on the wall. There is a nice interpretation of the fact
that $T_A$ is smaller than $T_B$. We can imagine that the vortex
in $B$ is composed by a vortex in $A$ and the wall rolled around
it. Thus the vortex in $B$ is more heavy because it has a wall
around.

\subsubsection{Crossing in nonabelian theories \label{crossna}}

\begin{figure}[h!t]
\begin{center}
\leavevmode \epsfxsize 8 cm \epsffile{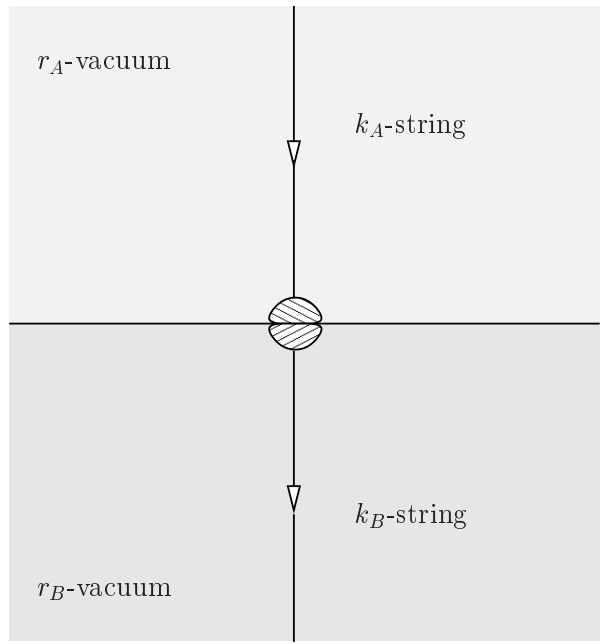}
\end{center}
\caption{\footnotesize The vortex crosses the wall. A bound state
confined on the wall can change the flux of the vortex. This
happens whenever $k_A-k_B$ is not a multiple of
$\gcd{(r_A-r_B,N)}$.} \label{crossSYM}
\end{figure}

The last example comes from $\N=1$ $SU(N)$ super Yang-Mills.  We
choose $r_A$, for the vacuum $A$, and $r_B$, for the vacuum $B$,
so that $r_B-r_A$ is a divisor of $N$. We use again the Rey's
argument to determine which string ends on the wall and which one
crosses the wall.  Is possible to build a bound state on which a
$k$-string could end, only if $k$ is a multiple of
$\gcd{(r_B-r_A,N)}$. Thus we have a subgroup
$\mathbb{Z}_{\gcd{(r_B-r_A,N)}} \subset \mathbb{Z}_N$ of strings
that can end on the wall with an End junction. Strings in the
quotient group $\mathbb{Z}_{N/\gcd{(r_B-r_A,N)}}$ instead, cross
the wall and goes into another string (Cross junction). Like in
Figure \ref{crossSYM} there can be a bound state multiple of
$\gcd{(r_B-r_A,N)}$ that change the flux of the string but cannot
absorb completely its flux.

We analyze the simplest example. Take $N$ even and a wall between
$r_A$ and $r_B=r_A+2$. Even strings can end on the wall. Odd
strings must cross the wall and they becomes odd strings on the
other side. For example a $1$-string can end on a bound state that
carries charge $Q^{-2}$ and then emerges on the other side like a
$-1$-string.

\subsection{A complete example \label{super}}

An interesting theory is $U(N)$ $\N=2$ super Yang-Mills broken to
$\N=1$ by a generic superpotential. This theory is enough rich to
present all the junctions just described.

The simplest example is a cubic superpotential so that
$W\p=(\Phi-a_1)(\Phi-a_2)$ has two roots. For simplification we
will consider the regime $\Lambda \gg a_1,a_2,a_1-a_2$. In this
case  the theory must distribute the eigenvalues of $\Phi$ at a
scale where the gauge group is still weakly coupled, and at this
scale, $U(N)$ is broken down to $U(N_1) \times U(N_2)$. At the
scale $\Lambda$ the gauge group confines and we have in global
$N_1 \cdot N_2$ vacua labelled by two integers $r_{N_1}$ and
$r_{N_2}$. A complete characterization of a vacua is thus given
specifying three integers $N_1$, $r_1$  and $r_2$. The confinement
index $t$ is the greatest common divisor between $N_1$, $N$, and
$r_1-r_2$ \cite{Cachazo:2002zk, Homotopy}. A domain wall is given
choosing a couple of vacua, call the first vacuum $A$ and the
second vacuum $B$.

\subsubsection{All the junctions simultaneously}

\begin{figure}[h!t]
\begin{center}
\leavevmode \epsfxsize 9 cm \epsffile{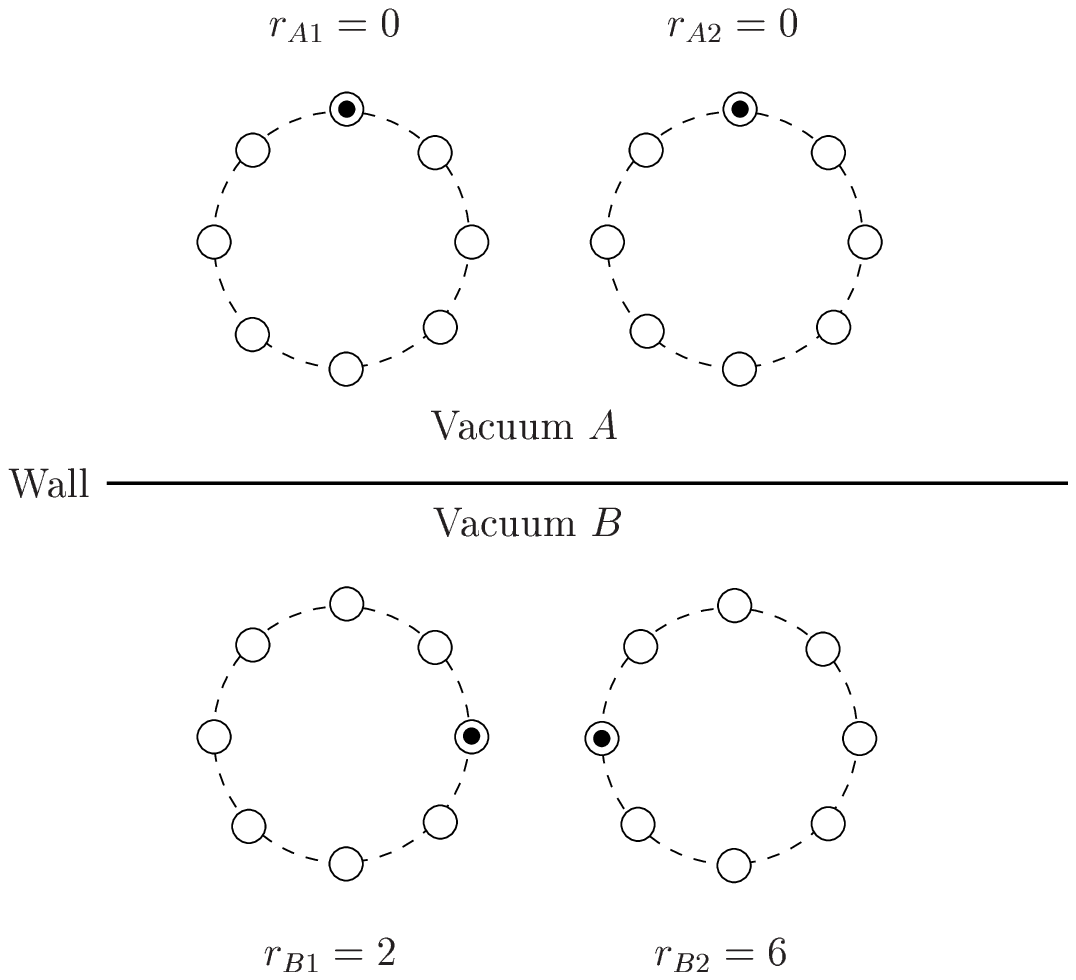}
\end{center}
\caption{\footnotesize $U(16) \to U(8) \times U(8)$.}
\label{tutti}
\end{figure}

The simplest case in which all the three kind of junctions appear
simultaneously is the one schematically represented if Figure
\ref{tutti}.
It is a $U(16)$ theory that breaks down to $U(8) \times U(8)$. For
vacuum $A$ we chose $r_{A 1}=r_{A 2}=0$ and the confinement index
is $t_A=8$. For vacuum $A$ we chose $r_{B 1}=2$, $r_{B 2}=6$ and
the confinement index is $t_B=4$. Vacuum $A$ has $\Z_8$ strings
while vacuum $B$ has $\Z_4$ strings. The state $Q^4$ is confined
in $A$ and non confined in $B$, thus we have a $\Z_2$ subgroup of
strings in $A$ that make a Coulomb junction with the wall. The
quotient $\Z_8 / \Z_2 =\Z_4$ is now equal to the strings in vacuum
$B$. The subgroup $\Z_2$ that confines even powers of $Q$ can end
on the wall and thus they form an End junction. The quotient $\Z_4
/ \Z_2 =\Z_2$ are the strings that confine odd powers of $Q$ and
they form a Cross junction.

\begin{figure}[h!t]
\begin{center}
\leavevmode \epsfxsize 6 cm \epsffile{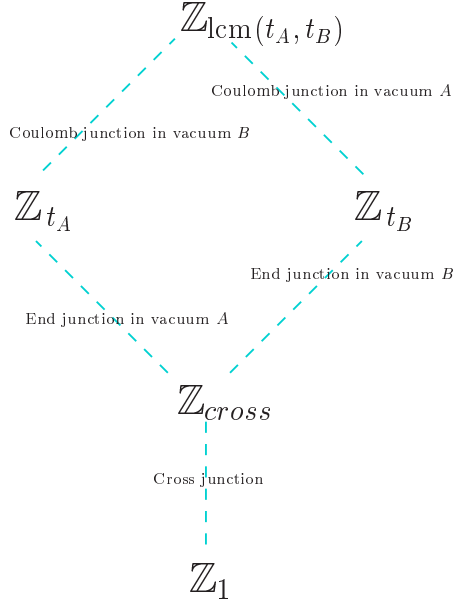}
\end{center}
\caption{\footnotesize  The web of groups that describe wall
vortex junctions between vacua of a $SU(N)$ gauge theory with
confinement index $t_A$ and $t_B$. The smallest group that
contains both the $A$ and the $B$ strings is
$\mathbb{Z}_{\lcm(t_A,t_B)}$. The subgroup
$\mathbb{Z}_{\lcm(t_A,t_B)/t_B}$ identifies strings in vacuum $A$
that end on the wall and forms a Coulomb junction. The opposite is
true for strings in vacuum $B$. $\mathbb{Z}_{cross}$ is the group
of stable strings in the domain wall background, that is the group
of strings that cross the wall. Strings in vacuum $A$ that belongs
to the subgroup $\mathbb{Z}_{t_A/ cross} \in \mathbb{Z}_{t_A}$ end
on the wall and form an end junction. The broup of crossing
strings $\mathbb{Z}_{cross}$ is the quotient
$\mathbb{Z}_{t_A}/\mathbb{Z}_{t_A/ cross}$.  The opposite is true
for strings in vacuum $B$. } \label{groups}
\end{figure}

\begin{figure}[h!t]
\begin{center}
\leavevmode \epsfxsize 14 cm \epsffile{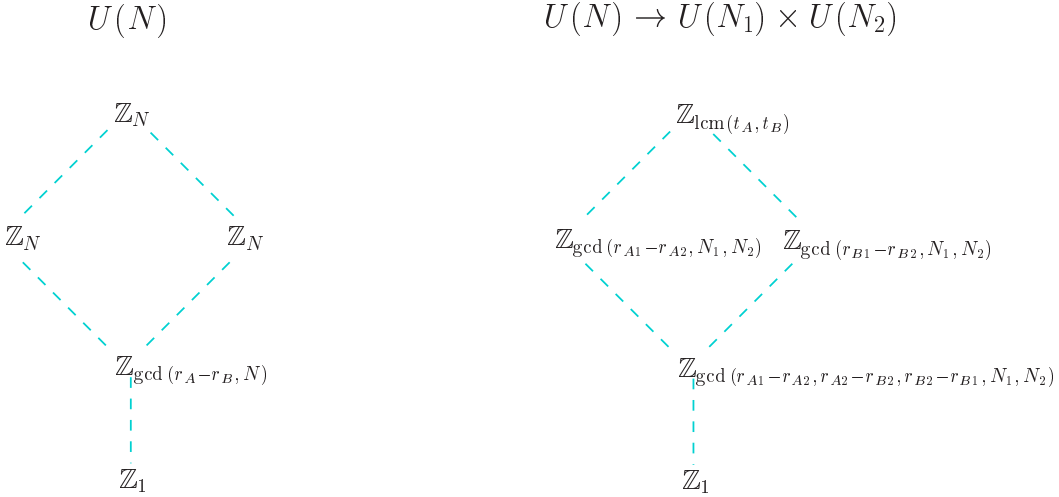}
\end{center}
\caption{\footnotesize  Two examples of web of groups in Figure
\ref{groups}. The first refers to $U(N)$ super Yang-Mills where
vacuum $A$ has index $r_A$ and vacuum $B$ has index $r_B$. The
second example is $U(N) \to U(N_1) \times U(N_2)$ with indices
$r_{A 1}$, $r_{A 2}$ for vacuum $A$ and $r_{B 1}$, $r_{B 2}$ for
vacuum $B$. } \label{esempi}
\end{figure}

In general all the three kind of junctions appear when we have
vacua with different confinement indices. What we are going to say
is presented schematically in Figure \ref{groups}.
The confining strings $\mathbb{Z}_{t_A}$ of vacuum $A$ and
$\mathbb{Z}_{t_B}$ of vacuum $B$, must be though embedded in the
smallest group that contains both of them
$\mathbb{Z}_{\lcm{(t_A,\,t_B)}}$. Strings that are multiple of
$t_B$ but not of $t_A$ are wall vortices in vacuum $B$ and they
belongs to the group $\mathbb{Z}_{\lcm{(t_A,\,t_B)}/t_B}$. In the
same way  strings that are multiple of $t_A$ but not of $t_B$ are
wall vortices in vacuum $A$ and they belongs to the group
$\mathbb{Z}_{\lcm{(t_A,\,t_B)}/t_A}$. For the remaining strings we
must decide if they form an End junction or a Cross junction. The
Cross-junctions has a nice interpretation. They can be thought as
the stable strings in the domain wall background. So in the same
way $\mathbb{Z}_{t_A}$ is the group of string in vacuum $A$ and
$\mathbb{Z}_{t_B}$ is the group of strings in vacuum $B$, so
$\mathbb{Z}_{cross}$ is the group of stable strings in the domain
wall background where $cross$ must be a divisor of both $t_A$ and
$t_B$. The remaining strings $\mathbb{Z}_{t_A/cross}$ are End
junction in vacuum $A$ and $\mathbb{Z}_{t_B/cross}$ are End
junctions in vacuum $B$. In Figure \ref{esempi} we give two
examples of the groups relation of the scheme in Figure
\ref{groups}.

\subsubsection{Decay of junctions}

In the theory at hand, strings can appear in different regimes
distinguished by the magnitude of the superpotential. $\sqrt{W\p}$
is the energy scale where $\N=2$ is broken and $W\p$ is the
tension scale of the strings. Thus we have three cases.
$\sqrt{W\p} \ll \Lambda \ll a$ is the strong coupling regime,
where the superpotential is a small perturbation that can be added
to the low energy dynamics of $\N=2$. In this case flux tubes are
nothing but ANO vortices of the low-energy $U(1)$'s. In the
intermediate regime $\Lambda \ll \sqrt{W\p} \ll a$, the dynamics
is approximatively that of various independent $U(N_i)$, note that
the number of strings is the same as before but now they are
governed by the group $\oplus_i \mathbb{Z}_{N_i}$. The transition
between the strong coupling regime and the pure $\N=1$ regime has
been studied in \cite{Hanany:1997hr} and there is nothing new to
say about it. The last regime is $\Lambda \ll a \ll \sqrt{W\p}$,
where some strings that where metastable before now can decay.
Their dynamics is governed by the group
$\mathbb{Z}_{\gcd{(N_i,b_i)}}$.

\begin{figure}[h!t]
\begin{center}
\leavevmode \epsfxsize 9 cm \epsffile{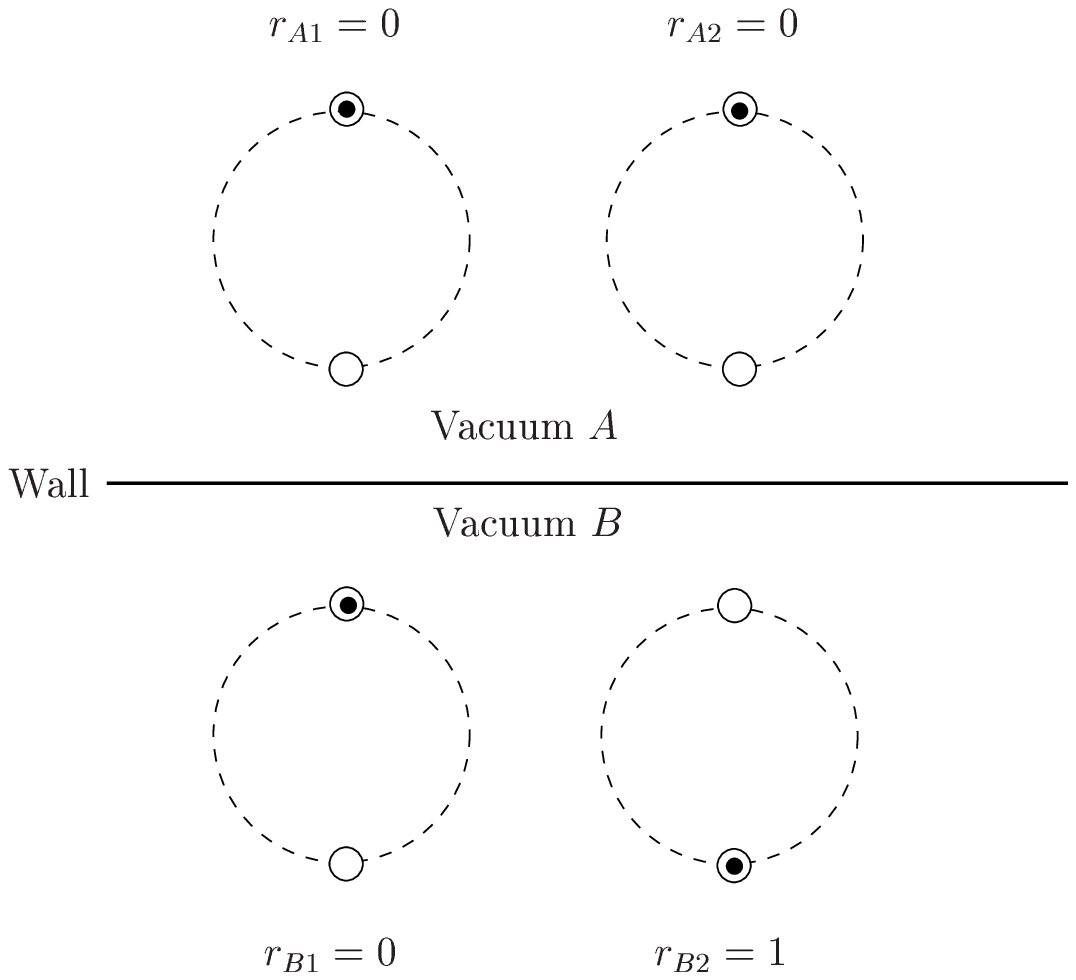}
\end{center}
\caption{\footnotesize $U(4) \to U(2) \times U(2)$.}
\label{decayth}
\end{figure}

\begin{figure}[h!t]
\begin{center}
\leavevmode \epsfxsize 10 cm \epsffile{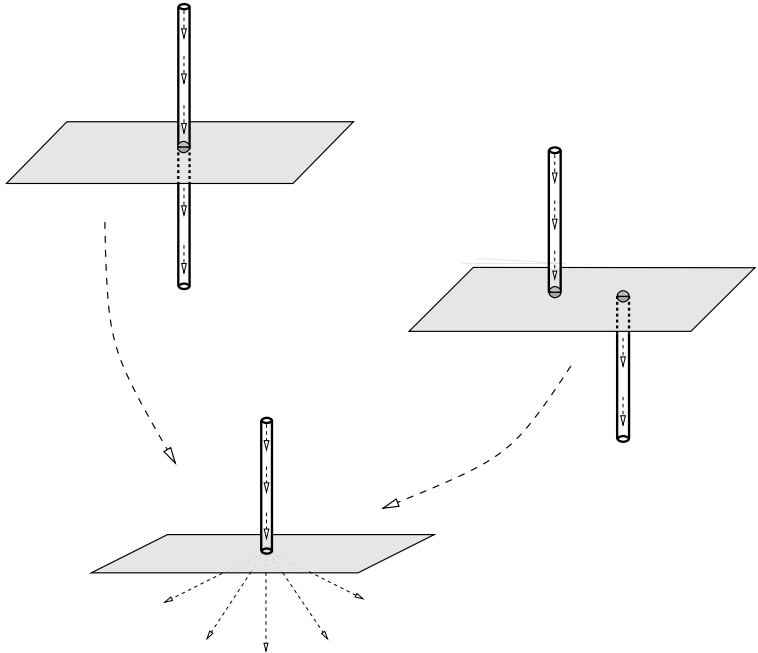}
\end{center}
\caption{\footnotesize Decay of junctions.} \label{decay}
\end{figure}

We consider in detail the case $N=4$. We choose both the vacua $A$
and $B$ so that classically $U(4) \to U(2) \times U(2)$. In vacuum
$A$ both the $U(2)$ factors are in the monopole vacuum $r_{A
1}=r_{A 2}=0$. In vacuum $B$ the first $U(2)$ is in the monopole
vacuum $r_{B 1}=0$, and the second in the dyon vacuum $r_{B 2}=1$
(see Figure \ref{decayth}).
\begin{description}
    \item[$\mathbf{\Lambda \ll \sqrt{W\p} \ll a}$.]    This is
the most easiest region to analyze. Both vacua have two confining
strings that are the non trivial elements of the group
$\mathbb{Z}_2 \oplus \mathbb{Z}_2$. The first group refers to the
first $U(2)$ and, since $A$ and $B$ are both in the monopole
vacua, these strings cannot end on the wall. So we are forced to
conclude that they form a Cross junction. The second
$\mathbb{Z}_2$ refers to the second $U(2)$. Since vacuum $A$ is in
the monopole vacuum and $B$ in the dyon vacuum, the strings can
end on the wall and so they forms an End junction.
\item[$\mathbf{\Lambda \ll a \ll \sqrt{W\p}}$.]   In this region
of parameters the two theories are governed by the confinement
index. Vacuum $A$ has $\mathbb{Z}_2$ strings and so only one non
trivial element. The interpretation is that the two elements
founded in the previous regime belongs to the same topological
element and can decay one into the other. Vacuum $B$ has $t_B=1$
so is completely unconfined. The two strings founded previously
are only metastable. The conclusion is obvious, we are in the
presence of a Coulomb junction. Thus we have seen an example of
the decay schematically presented in Figure \ref{decay}.
\end{description}

\appendix

\section{The Flux Tube/Flux Tube Junction} \label{yappendix}

In this Appendix we describe, using the things studied in this
paper, a way to obtain a  junction between a flux tube and another
flux tube. This junction can be used to build a baryon vertex in
some particular confining gauge theory. In the string model of
hadrons, mesons are identified with a string with a quark and an
antiquark attached at the endpoints. For the baryons things are
not so natural and various configurations are possible. One of the
hypothesis is the so called   ``Y'' junction: the baryon is
composed by three strings with one end in common and three quarks
at the other ends \cite{Artru:1974zn}. In general the construction
of the baryon vertex is not at all trivial (see for example
\cite{Hanany:1997hr} for  MQCD and \cite{Witten:1998xy} for the
AdS/CFT correspondence).      In what follows we describe another
possible mechanism for the formation of the Y junction using the
flux tube/flux tube junction.

 Suppose to have a theory with two gauge
groups $U(1)_1$ and $U(1)_2$, and two discrete vacua $A$ and $B$.
For the first $U(1)_1$ both vacua are in the Higgs phase and, on
the domain wall interpolating between the two, there is a
localization of the gauge field. Thus we have a wall/vortex
junction like a D-brane. For the second group $U(1)_2$ the vacuum
$A$ is in the Higgs phase, while vacuum $B$ is in the Coulomb
phase and we can have a vortex made of wall like the one studied
in Section \ref{made}. The vortex in vacuum $A$ that carries the
magnetic flux of $U(1)_1$, can end on the domain wall and, as a
consequence, can also end on the magnetic vortex that carries flux
under $U(1)_2$. In fact the last one is nothing but the wall
rolled in a cylinder.  In this way we we have obtained a ``Y''
junction.

Now we are going to show take a particular confining $SU(3)$ gauge
theory that admits the  Y junction previously constructed as its
baryon vertex.

The theory that we consider is $\N=2$ $SU(3)$ SQCD with two
hypermultiplets in the foundamental representation. the hypermultiplets have mass $m_A$ and $m_B$. In a
generic point of the moduli space ($2$ complex dimensions) the
theory is in a $U(1)_1 \times U(1)_2$ Coulomb phase. At some
critical lines ($1$ complex dimension), there are massles
hypermultiplets for one of the $U(1)$'s. At some crtical points ($0$
complex dimensions), both the U(1)'s have charged massless
hypermultiplets. We take vacuum $A$ to be the critical point where
$U(1)_1$ is locked to the flavor $m_A$, and some massless
monopole is charged under $U(1)_2$. We choose vacuum $B$
to be on the critical line where $U(1)_1$ is locked to $m_2$, while
$U(1)_2$ has no massless charged particles.  We know that is
possible to choose some superpotential $W(\Phi)$ so that vacua $A$
and $B$ are among the ones that survive the perturbation. After
the perturbation, charged  massless particles condense and creates flux
tubes. If these particles have magnetic charge,  the flux tubes carry electric flux and are responsible for confinement.

It is possible to send the mass $m_1$  into the strong coupling region ($\sim \Lambda$),
 so that  the charge of the locked massless flavor becomes magnetic, and vacuum A is in the confining phase.

\begin{figure}[h!t]
\begin{center}
\leavevmode \epsfxsize 10 cm \epsffile{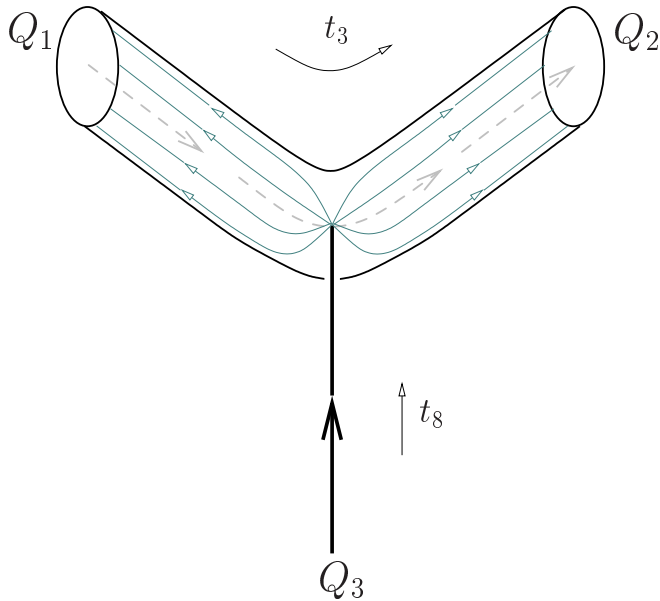}
\end{center}
\caption{\footnotesize A Y junction for the baryon vertex. The
three quarks are choosen so that the foundamental of $SU(3)$ is
$(Q_1,Q_2,Q_3)$. $Q_1$ and $Q_2$ are connected by a wall vortex
that carries the flux of the $t_3$ generator. The quark $Q_3$ is
connected to another tube that carries the $t_8$ charge makes a
junction like \ref{nahiggs} with the wall vortex. The charge of
$Q_3$ is divided in two equal parts when the $t_8$ tube join the
$t_3$ wall vortex, and this is consistent with the fact that $t_8
\propto \mathrm{diag}(1,1,-2)$. } \label{yjunction}
\end{figure}

With respect to $U(1)_1$ both vacua are both in the Higgs phase
but the breaking of the gauge group is due to two different
charged particles. Thanks to the same mechanism of
\cite{Shifman:2002jm}, the $U(1)_1$ gauge field is localized on the wall that interpolates between $A$ and $B$.
 With respect to $U(1)_2$, vacuum $A$ is in
the Higgs phase while vacuum $B$ in the Coulomb phase. The two flux tubes in vacuum $A$ can thus form a Y junction like the one previously described.
So we can
have a baryon vertex like the one of Figure \ref{yjunction}.







\section* {Acknowledgement}
 I want to thank the following people for useful comments and
 discussions: J.~Evslin, K.~Konishi, H.~B.~Nielsen, A.~Ritz and A.~Smilga.

\end{document}